\documentclass[preprint,amsmath,amssymb,aps,prd,nofootinbib]{revtex4}
\usepackage[utf8]{inputenc}
\pdfoutput=1 
\usepackage{graphics,graphicx}
\usepackage[usenames,dvipsnames]{color}

\def\bma{\left( \begin{array} }
\def\ema{\end{array} \right)}
\newcommand{\vdel}{v_{\Delta}}
\newcommand{\delm}{\Delta M}

\begin{document}

\preprint{\hfill{KIAS-P13054}}

\title{Search for a doubly-charged boson in
four lepton final states in type II seesaw}
\author{Eung Jin Chun and Pankaj Sharma}
\affiliation{
Korea Institute for Advanced Study, Seoul 130-722, Korea}

\begin{abstract}
CMS and ATLAS have searched for a doubly-charged boson $H^{\pm\pm}$
which may arise from type II seesaw in the 7 TeV run at the LHC
by considering pair or associated production of doubly-charged bosons
under the assumption of degenerate triplet scalars.
In this work, we consider non-degenerate triplet components with the mass gap
$\Delta M \sim 1 - 40$ GeV which leads to  enhanced pair-production cross-sections
of $H^{\pm\pm}$ added by the gauge decays of the heavier neutral and singly-charged bosons.
We reevaluate the constraints in the $\Delta M-M_{H^{++}}$ plane depending on the triplet
vacuum expectation value $v_\Delta$ in the type II seesaw model
which are much more stringent than the current search limits.
We further study the possibility of observing same-sign tetra-lepton signals
in the allowed parameter space which can be probed in the future runs of the LHC.
\end{abstract}

\maketitle

\section{Introduction}
One of the key questions in physics beyond Standard Model is the origin of the neutrino masses
and mixing. It can be attributed to an $SU(2)$ triplet boson which couples
to both the lepton doublet fermions and the Higgs doublet boson  realizing the
so-called type II seesaw mechanism \cite{type2}. An essential feature of this scenario is the
presence of a doubly-charged boson $H^{\pm\pm}$
whose decay to same-sign di-leptons with different flavor
states may allow us to probe the neutrino mass structure at the LHC \cite{chun03}.
CMS \cite{Chatrchyan:2012ya} and ATLAS \cite{ATLAS:2012hi} have searched for doubly-charged
bosons at $\sqrt{s}=7$ TeV with about 5 fb$^{-1}$ of integrated luminosity of data.
CMS have considered three- and four-lepton final states coming from
the associated production process $pp\to H^{++}H^{-}\to\ell^+_i\ell^+_j\ell^-_k\nu_l$ \cite{Akeroyd05} 
and the pair production process
$pp\to H^{++}H^{--}\to\ell^+_i\ell^+_j\ell^-_k\ell^-_l$ \cite{Huitu96}
to put constraints on the doubly-charged boson mass $M_{H^{++}}$ in four different
benchmark points that would probe different neutrino mass structure.
On the other hand, ATLAS looked at same-sign di-lepton (SS2L) signals to probe $H^{\pm\pm}$ in pair production of doubly-charged boson at the LHC.
In their analysis, they put strong bound on leptonic branching fractions of the doubly-charged boson depending on its mass.

In both analyses, degenerate masses for the triplet bosons, $H^{\pm\pm}$, $H^\pm$, $H^0$ and $A^0$
are assumed, which is possible only when a particular
scalar coupling called $\lambda_5$  in the scalar potential vanishes.
But there is no reason to assume this particular coupling to be zero. Indeed, interesting
phenomena arise for non-vanishing $\lambda_5$ \cite{chun03,Akeroyd11,Melfo11,Aoki11}.
When $\lambda_5$ is positive leading to $\Delta M \equiv M_{H^+} - M_{H^{++}} \approx M_{H,A}-M_{H^+} >0$, 
$H^{\pm\pm}$ is the lightest among triplet scalars and other triplet scalars decay
dominantly to $H^{\pm\pm}$ through cascade decay
associated with several $W^{\mp*}$
 in a large parameter space of $\lambda_5$.\footnote{The mass gap $\Delta M$ is restricted by
$|\Delta M| \lesssim 40$ GeV independently of $M_{H^{++}}$ due to electroweak precision
constraints \cite{Chun:2012jw} and thus the associated $W^\pm$ are always off-shell.} 
In this parameter space, pair-production cross section is
enhanced significantly since other (pair and associated) triplet production channels contribute to
pair-production of doubly-charged bosons. This leads to a more stringent bound on doubly-charged boson mass
 $M_{H^{++}}$ as compared to the current CMS and ATLAS bounds.

In this paper, we evaluate the exclusion regions in the $M_{H^{++}}$--$\Delta M$ plane 
in the type II seesaw model utilizing the search strategy employed by CMS and ATLAS collaborations.
We consider $\lambda_5$ (and thus $\Delta M$) to be non-vanishing and thus expect much stronger bound on 
$M_{H^{++}}$ than obtained by CMS and ATLAS. This bound depends also on the triplet
vacuum expectation value $v_\Delta$ which controls the ratio of the branching fractions
for $H^{++} \to l^+_{i} l^+_j$ and $W^+ W^+$ through the neutrino mass
relation \cite{chun03}.
 For the illustration of our analysis, we choose
three different values of $v_\Delta$ to examine the parameter regions
of $(M_{H^{++}}, \Delta M)$ allowed by the current data and then
look for the possibility of observing same-sign tetra-lepton (SS4L) signal \cite{Chun:2012zu} at 8 TeV LHC (LHC8),
and 13 TeV LHC (LHC13) with 20 fb $^{-1}$ and 100 fb$^{-1}$ integrated luminosities, respectively. 
When $v_\Delta \gg 10^{-4}$ GeV, the branching fraction of $H^{++} \to W^+ W^+$ is almost 100 \% resulting in
highly suppressed same-sign di-lepton \cite{Han07} or four lepton signals \cite{Kanemura13}
from $W$ decays and thus very loose bounds on $M_{H^{++}}$. 
We take $v_\Delta$ as large as $2 \times 10^{-4}$ GeV for which the 
branching fraction of $H^{++} \to l^+ l^+$ is around 20 \% and thus still a sizable number of four lepton final stastes
can arise.
Note that SS4L signals arise due to a novel phenomenon of the triplet-antitriplet oscillation
guaranteed by a tiny mass splitting between $H^0$ and $A^0$ related to the neutrino mass,
which leads to pair-production of same-sign doubly-charged bosons after the chain decays of
$H^0,A^0 \to H^{\pm} \to H^{\pm\pm}$ allowed by sizable $\Delta M$ \cite{Chun:2012zu}.

\section{Type II Seesaw Model}

When the Higgs sector of the Standard Model is extended to have a
$Y=1$ complex $SU(2)_L$ scalar triplet $\Delta$ in addition to the standard doublet $\Phi$,
the gauge-invariant Lagrangian is written
as
\begin{eqnarray}
\mathcal L=\nonumber\left(D_\mu\Phi\right)^\dagger
\left(D^\mu\Phi\right)  + \mbox{Tr}
\left(D_\mu\Delta\right)^\dagger\left(D^\mu\Delta\right) -\mathcal
L_Y - V(\Phi,\Delta)
\end{eqnarray}
where the leptonic part of the Lagrangian required to generate
neutrino masses is
\begin{equation} \label{leptonYuk}
\mathcal L_Y= f_{ij}L_i^T Ci\tau_2\Delta L_{j} +
\mbox{H.c.}
\end{equation}
and the scalar potential is
\begin{eqnarray}\label{Pot}
V(\Phi,\Delta)&=&\nonumber m^2\Phi^\dagger\Phi +
\lambda_1(\Phi^\dagger\Phi)^2+M^2\mbox{Tr}(\Delta^\dagger\Delta)\\\nonumber
&+&\lambda_2\left[\mbox{Tr}(\Delta^\dagger\Delta)\right]^2+\lambda_3\mbox{Det}(\Delta^\dagger\Delta)
+\lambda_4(\Phi^\dagger\Phi)\mbox{Tr}(\Delta^\dagger\Delta)\\
&+&\lambda_5(\Phi^\dagger\tau_i\Phi)\mbox{Tr}(\Delta^\dagger\tau_i\Delta)
+\left[\frac{1}{\sqrt{2}}\mu(\Phi^Ti\tau_2\Delta\Phi)+\mbox{H.c.}\right].
\end{eqnarray}
Here used is the $2\times 2$ matrix representation of $\Delta$:
\begin{equation}
\Delta=\bma{cc}
\Delta^+/\sqrt{2}  & \Delta^{++} \\
\Delta^0       & -\Delta^+/\sqrt{2} \ema .
\end{equation}
Upon the electroweak symmetry breaking with $\langle
\Phi^0\rangle=v_0/\sqrt{2}$, the $\mu$ term in Eq.~(\ref{Pot}) gives rise to the
vacuum expectation value of the triplet $\langle
\Delta^0\rangle=v_\Delta/\sqrt{2}$ where $\vdel\approx \mu
v_0^2/\sqrt{2}M^2$. For non-vanishing $v_\Delta$,
the neutrino mass matrix is generated as a product of the leptonic Yukawa
coupling (\ref{leptonYuk}) and $v_\Delta$:
\begin{equation} \label{Mnu}
 M^\nu_{ij} = f_{ij} v_\Delta \,.
\end{equation}
This allows us to reconstruct the Yukawa matrix $f_{ij}$
from the current neutrino oscillation data up to unmeasured CP phases and mass hierarchy.
For our analysis, we use two neutrino mass matrices for normal and inverted
hierarchies derived in Ref.~\cite{Chun:2012zu} assuming vanishing CP phases.

After the electroweak symmetry breaking, there are seven physical
massive scalar eigenstates denoted by $H^{\pm,\pm}$, $H^\pm$,
$H^0$, $A^0$, $h^0$. Under the condition that $|\xi|\ll 1$ where
$\xi \equiv v_\Delta/v_0$, the first five states are mainly from
the triplet scalar and the last from the doublet scalar. For the
neutral pseudoscalar and charged scalar parts,
\begin{eqnarray}
 \phi^0_I = G^0 - 2 \xi A^0 \;, \qquad
 \phi^+ = G^+ + \sqrt{2} \xi H^+  \nonumber\\
 \Delta^0_I= A^0 + 2 \xi G^0 \;, \qquad
 \Delta^+= H^+ - \sqrt{2} \xi G^+
\end{eqnarray}
where $G^0$ and $G^+$ are the Goldstone modes, and
for the neutral scalar part,
\begin{eqnarray}
 \phi^0_R &=& h^0 - a \xi \, H^0 \,, \nonumber\\
 \Delta^0_R&=& H^0 + a \xi \, h^0
\end{eqnarray}
where $ a = 2 + 4 (4\lambda_1-\lambda_4-\lambda_5) M^2_{W}/
   g^2(M^2_{H^0}-M^2_{h^0}) $.
Neglecting the triplet--doublet mixing, the masses of the triplet bosons are
\begin{eqnarray} \label{massD}
 M^2_{H^{\pm\pm}} &=& M^2 + 2{\lambda_4 -\lambda_5 \over g^2 } M^2_{W}
 \nonumber\\
 M^2_{H^{\pm}} &=& M_{H^{\pm\pm}}^2 + 2{\lambda_5 \over g^2} M^2_{W}
 \nonumber\\
 M^2_{H^0, A^0} &=&  M^2_{H^\pm} +
    2{\lambda_5 \over g^2} M^2_{W} \,.
\end{eqnarray}
The mass of the Standard Model boson $h^0$ is
given by $m_{h^0}^2=4\lambda_1 v_\Phi^2$ as usual.

Eq.~(\ref{massD}) tells us that the mass splitting among triplet
scalars to the linear order for small splitting (that is, for
$|\lambda_5| M_W\ll gM$) can be written as
\begin{equation}
 \delm \approx \frac{\lambda_5 M_W}{g}.
\end{equation}
Furthermore,  depending upon the sign of the coupling $\lambda_5$,
there are two mass hierarchies among the triplet components:
$M_{H^{\pm\pm}}>M_{H^\pm}>M_{H^0/A^0}$ for $\lambda_5<0$; or
$M_{H^{\pm\pm}}<M_{H^\pm}<M_{H^0/A^0}$ for $\lambda_5> 0$. In this
work, we focus on the latter scenario,  where the doubly-charged
scalar $H^{\pm\pm}$ is the lightest so that it decays only to
$l_i^\pm l_j^{\pm}$ or $W^\pm W^\pm$ whose coupling
constants are proportional to $f_ij$ or $\xi$,
respectively:
\begin{eqnarray} \label{ll-WW}
{\cal L} &=& {1\over\sqrt{2}} \left[ f_{ij} \bar l^c_i P_L l_j
+ g \xi M_W W^- W^-\right] H^{++} + h.c.
\end{eqnarray}
 Thus the branching fraction for $H^{++} \to l^+_i l^+_j$ is
completely determined for given $v_\Delta$ and the neutrino matrix (\ref{Mnu}).
 On the other hand, $H^0/A^0$  ($H^\pm$) decays
mainly to $H^\pm W^{\mp *}$ ($H^{\pm\pm} W^{\mp *}$) unless the
mass splitting $\Delta M$ is negligibly small.

\begin{figure}[t]
\begin{center}
\includegraphics[width=80mm]{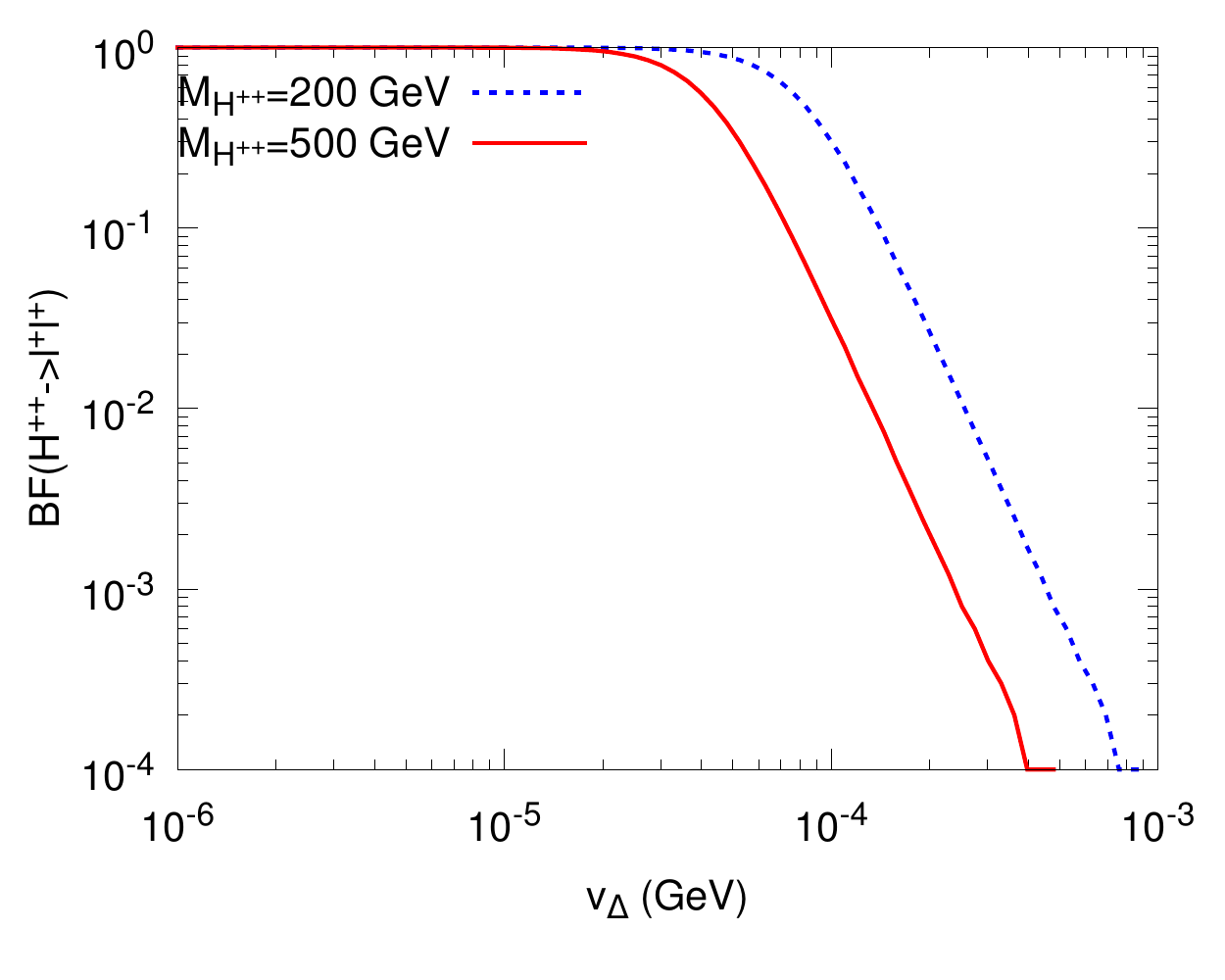}
\end{center}
\caption{ Branching fraction of $H^{++}\to \ell^+\ell^+$ as a function of triplet vacuum expectation value $v_\Delta$
for $M_{H^{++}}=200$ GeV and 500 GeV.
\label{BF-Hpp}}
\end{figure}

The di-lepton decay rates of $H^{++}$ are given by
\begin{equation}
\Gamma_{l_i l_j} \equiv  \Gamma(\Delta^{++} \to l^+_i l^+_j) = S {|f_{ij}|^2 \over 16\pi} M_{\Delta^{++}}
 \end{equation}
where $S=2\, (1)$ for $i\neq j\, (i=j)$. From the neutrino mass relation,
$M^\nu_{ij} = f_{ij} v_\Delta$, one gets the total di-lepton rate which is inversely proportional to $v_\Delta^2$:
\begin{equation}
\Gamma_{ll} \equiv \sum_{i,j} \Gamma_{l_i l_j} =  {1\over 16\pi} {{\bar m_\nu}^2 \over  v_\Delta^2} M_{\Delta^{++}}
\end{equation}
where $\bar m_\nu^2 = \sum_i m_{\nu_i}^2$ is the sum of three neutrino mass-squared eigenvalues.
On the other hand, the di-$W$ decay rate $\Gamma_{WW} = \Gamma(H^{++} \to W^{+} W^{+})$ 
is proportional to $v_\Delta^2$, and thus the leptonic branching fraction 
$\mbox{BF}(H^{++} \to l^+ l^+)\equiv \Gamma_{ll} / \Gamma_{H^{++}}$ is a sensitive function of $v_\Delta$. 
In Fig.~\ref{BF-Hpp}, we provide a plot for the leptonic branching fraction
depending on $v_\Delta$ for two values of $M_{H^{++}} = 200$ and 500 GeV.
For our collider analysis in the following sections, we will take three example values of $v_\Delta$ to discuss the dependence on
the leptonic branching fraction and the mass gap.

Given the neutrino mass matrices for the normal (NH)  and
inverted (IH) hierarchies \cite{Chun:2012zu},  the individual di-lepton decay rates
$\Gamma_{l_i l_j}$  normalized by the total leptonic decay rate $\Gamma_{ll}$ are given by
\begin{equation} \label{BRij}
\begin{tabular}{|c|c|c|c|c|c|c|}
\hline
$\Gamma_{l_i l_j}/\Gamma_{ll}$ (\%) &  $ee$ & $e\mu$  & $e\tau$ & $\mu\mu$  & $\mu\tau$ & $\tau\tau$ \\
\hline
NH & ~0.62~ & ~5.11~ & ~0.51~ & ~26.8~  & ~35.6~ & ~31.4~ \\
\hline
IH
 & 47.1 & 1.27 & 1.35 & 11.7  & 23.7 & 14.9 \\
\hline
\end{tabular}
\end{equation}
For given $v_\Delta$ one can read off the flavor-dependent branching fraction BF$(H^{++} \to l_i^+ l_j^+) = \Gamma_{l_i l_j} / 
\Gamma_{H^{++}}$ combining Eq.~(\ref{BRij}) and Fig.~\ref{BF-Hpp}. 

\medskip

An important quantity for a SS4L signal is the
mass splitting $\delta M_{HA}$ between $H^0$ and $A^0$ which is much smaller than the mass difference
$\Delta M$ between different triplet components. The $\mu$ term in Eq.~(\ref{Pot}),
which is lepton number violating, generates not only the triplet VEV:
\begin{equation}
 v_{\Delta} = {\mu  v_0^2 \over \sqrt{2} M^2_{H^0}}\,,
\end{equation}
but also the mass splitting between the heavy neutral scalars,
$\delta M_{HA} \equiv M_{H^0} - M_{A^0}$:
\begin{equation} \label{MHA}
 \delta M_{HA} = 2 M_{H^0} {v_\Delta^2\over v_0^2}
 { M^2_{H^0} \over M^2_{H^0} - m^2_{h^0} } \,.
\end{equation}
As will be shown later, $\delta M_{HA}$ can be comparable to the total decay rate
of the neutral scalars, $\Gamma_{H^0/A^0}$, for a preferable choice of $v_\Delta$,
which enhances the same-sign tetra lepton signal \cite{Chun:2012zu}.

\section{Constraining $\Delta M - M_{H^{\pm\pm}}$ from SS2L signals}

As stated earlier, CMS and ATLAS both have assumed degenerate triplet scalars and thus could only study
the process $pp\to H^{++}H^{--}$ in their analyses. In type II seesaw model, when scalar coupling $\lambda_5>0$,
there are several triplet scalar production processes which can contribute to pair-production of
doubly-charged bosons which are listed below:

\begin{enumerate}\label{proc}
\item $pp\to H^{\pm\pm}H^{\mp}$ followed by $H^{\mp}\to H^{\mp\mp}W^{\pm*}$,
\item $pp\to H^{\pm}H^{\mp}$ followed by $H^{\mp}\to H^{\mp\mp}W^{\pm*}$,
\item $pp\to H^{\pm}H^0/A^0$ followed by $H^0/A^0\to H^\mp W^{\pm*}$ and $H^{\pm}\to H^{\pm\pm}W^{\mp*}$.
\end{enumerate}

\begin{figure}[t]
\begin{center}
\includegraphics[width=80mm]{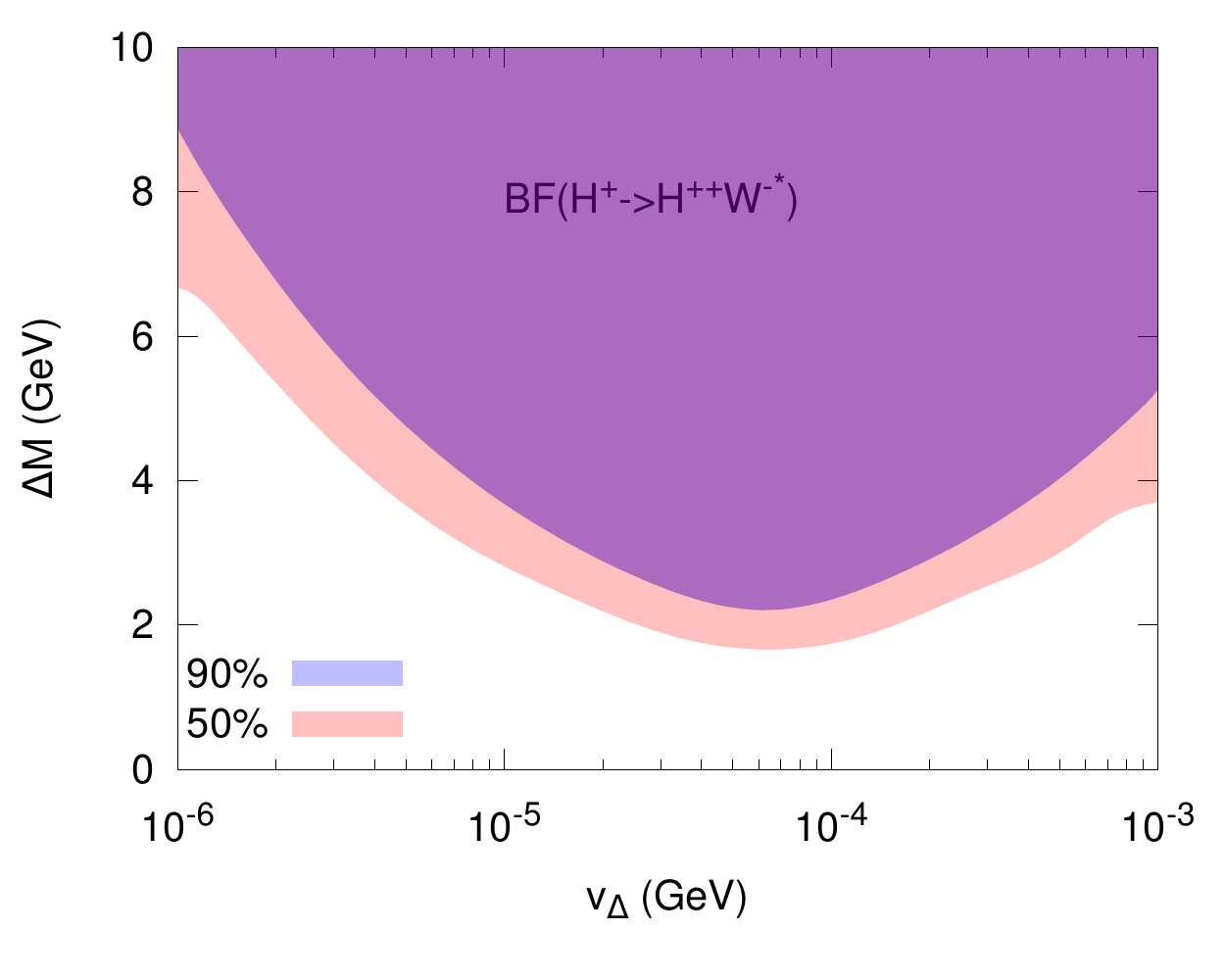}
\end{center}
\caption{Branching fraction of $H^\pm\to H^{\pm\pm}W^{\mp*}$ in the $v_\Delta-\Delta M$ plane
  for $M_{H^{++}}=$ 300 GeV.
 The purple and pink regions
denote the parameter space where BF($H^\pm\to H^{\pm\pm}W^{\mp*}$) is greater than 90\% and 50\% respectively.\label{BF-Hp}}
\end{figure}

In Fig.~\ref{BF-Hp}, we plot the branching fraction of $H^\pm\to H^{\pm\pm}W^{\mp*}$
in the $v_\Delta-\Delta M$ plane. The purple and pink regions denote the parameter space where
BF($H^\pm\to H^{\pm\pm}W^{\mp*}$) is greater than 90\% and 50\% respectively.
As can be seen from the figure that when there is non-zero mass splitting among
the triplet scalars, there can be large parameter space where this BF is dominant.
Furthermore this would lead to a significant enhancement in the number of events for pair-production of doubly-charged bosons and thus may result
in a more stringent constraints on doubly-charged boson mass $M_{H^{++}}$.

One of our aims in this paper is to
revise the constraints on $M_{H^{++}}$ obtained by CMS and ATLAS
after including all the processes which contribute to pair-production of doubly-charged bosons.
We use \texttt{CTEQ6L} \cite{Pumplin:2005rh} parton  distribution
function (PDF) and the renormalization/factorization scale is set
at $2M_{H^+}$. \texttt{CALCHEP} \cite{Pukhov:2004ca} is used to
generate the parton level events for the relevant processes. Then,
using \texttt{LHEF} \cite{Alwall:2006yp} interface, we pass these
parton level events to \texttt{PYTHIA} \cite{Sjostrand:2006za} for
fragmentation and initial/final state radiations. We use
\texttt{PYCELL}, a toy calorimeter in \texttt{PYTHIA}, for
hadronic level simulation for finding jets using a cone algorithm.
For a more realistic simulation, we utilize the same analysis strategy as
employed by CMS and ATLAS collaborations \cite{Chatrchyan:2012ya, ATLAS:2012hi}
in the study of doubly-charged boson. We use selection criteria for four lepton events
from table 3 of the CMS paper \cite{Chatrchyan:2012ya}.
As for the same-sign dilepton analysis which was performed by ATLAS, we put following
selection criteria. Leptons must have a transverse
momentum above 20 GeV and be well isolated. In pairs where the higher-$p_T$ lepton is
an electron, it is required to have $p_T > 25$ GeV. All pairs of
electrons or muons with the same electric charge are considered.
The invariant mass of the lepton pair must be larger
than 15 GeV, and for $e^\pm e^\pm$ the region close to the Z-boson
mass $(70~ \mbox{GeV} < m(e^\pm e^\pm) < 110~ \mbox{GeV})$ is excluded due to
a large background from $Z \to e^+e^-$ events with an electron
charge misidentification.

\begin{figure}[t]
\begin{center}
\includegraphics[width=80mm]{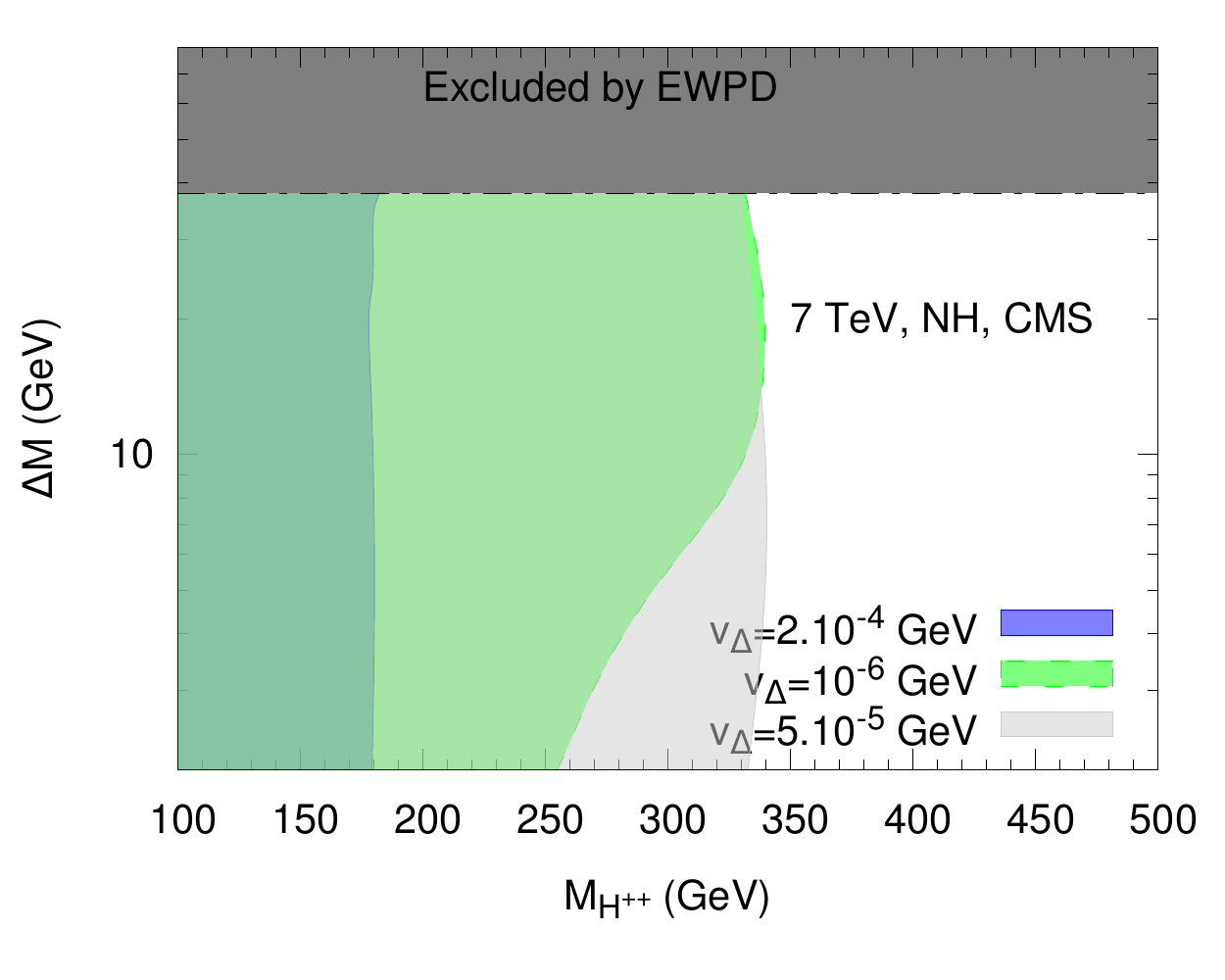}
\includegraphics[width=80mm]{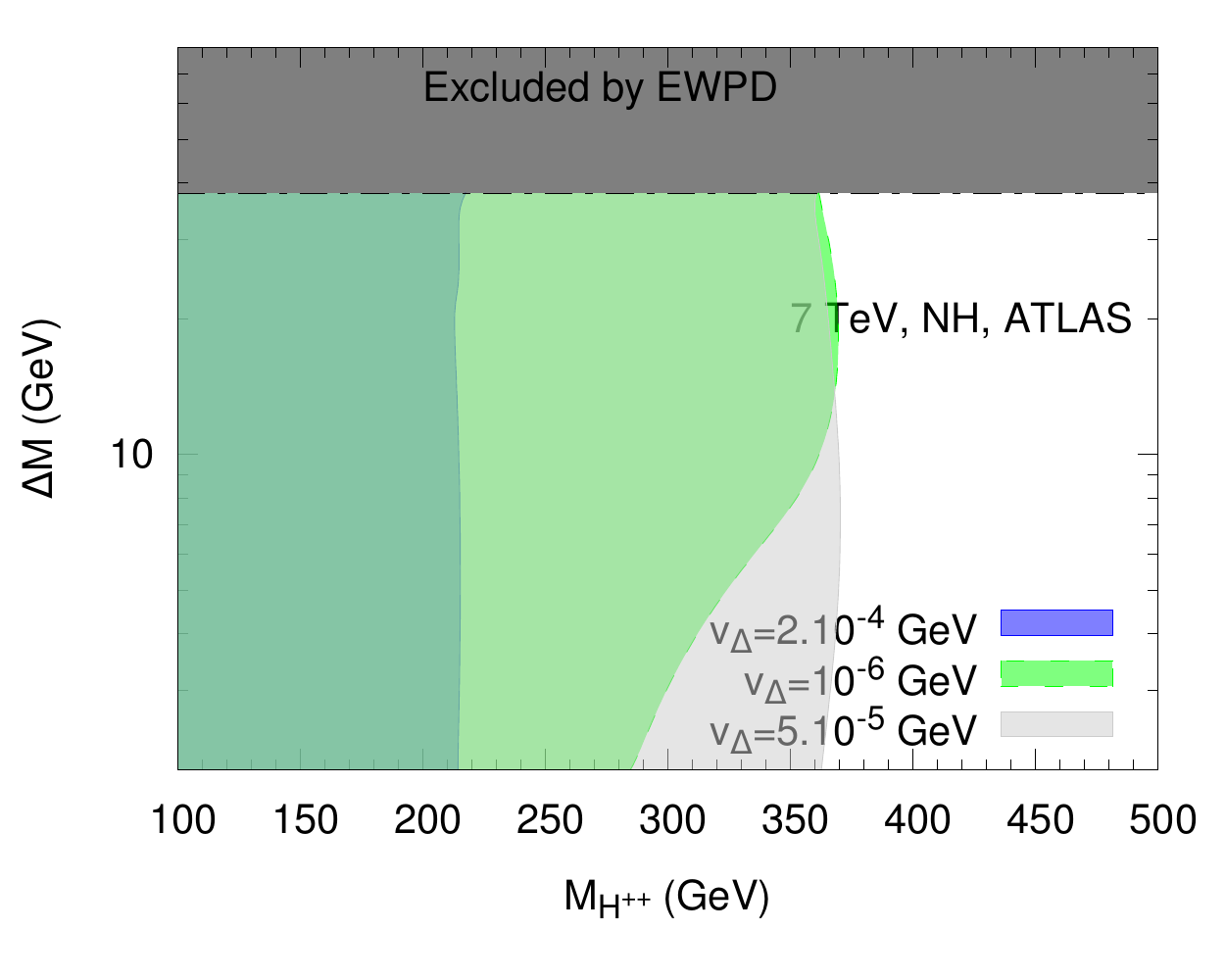}
\includegraphics[width=80mm]{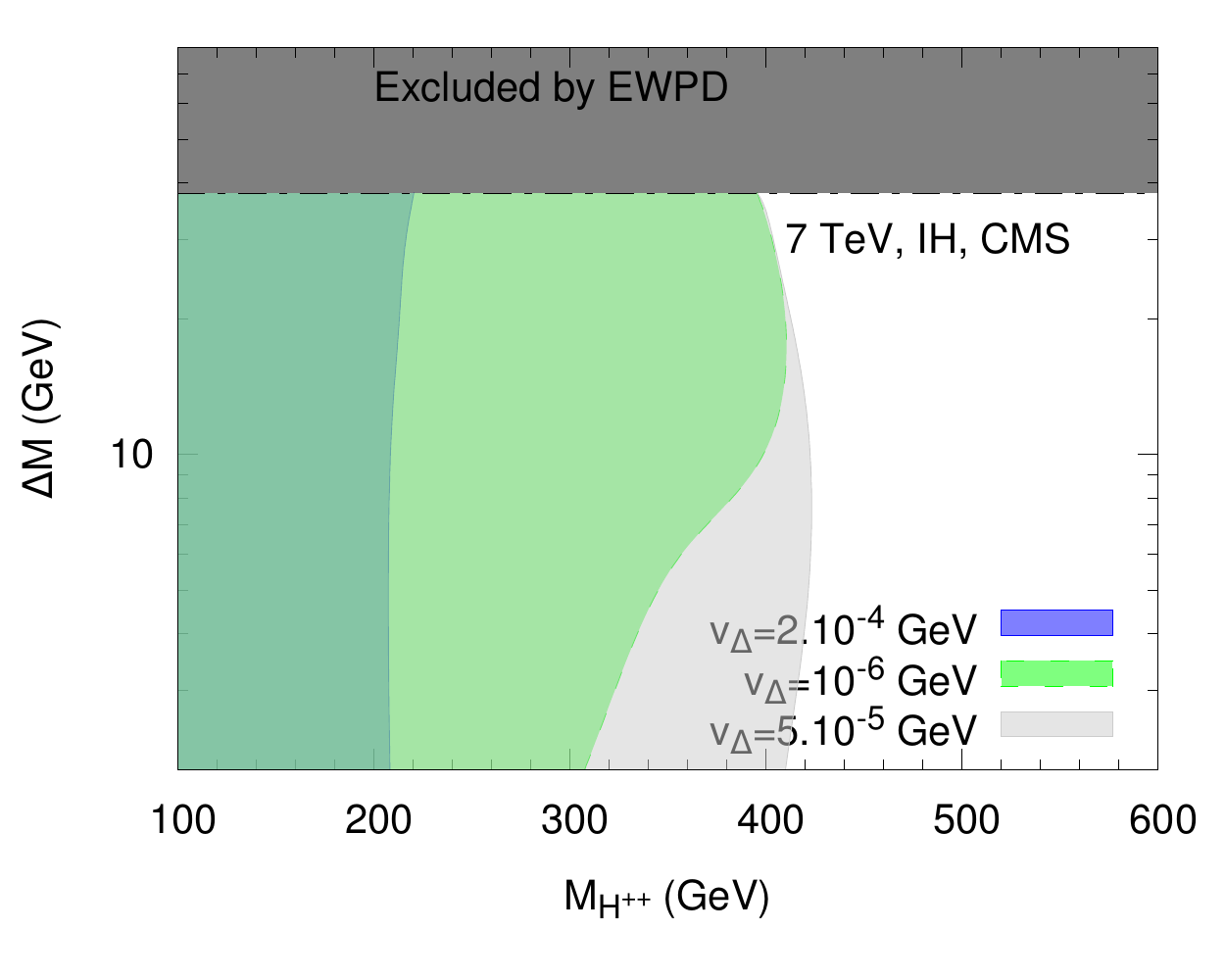}
\includegraphics[width=80mm]{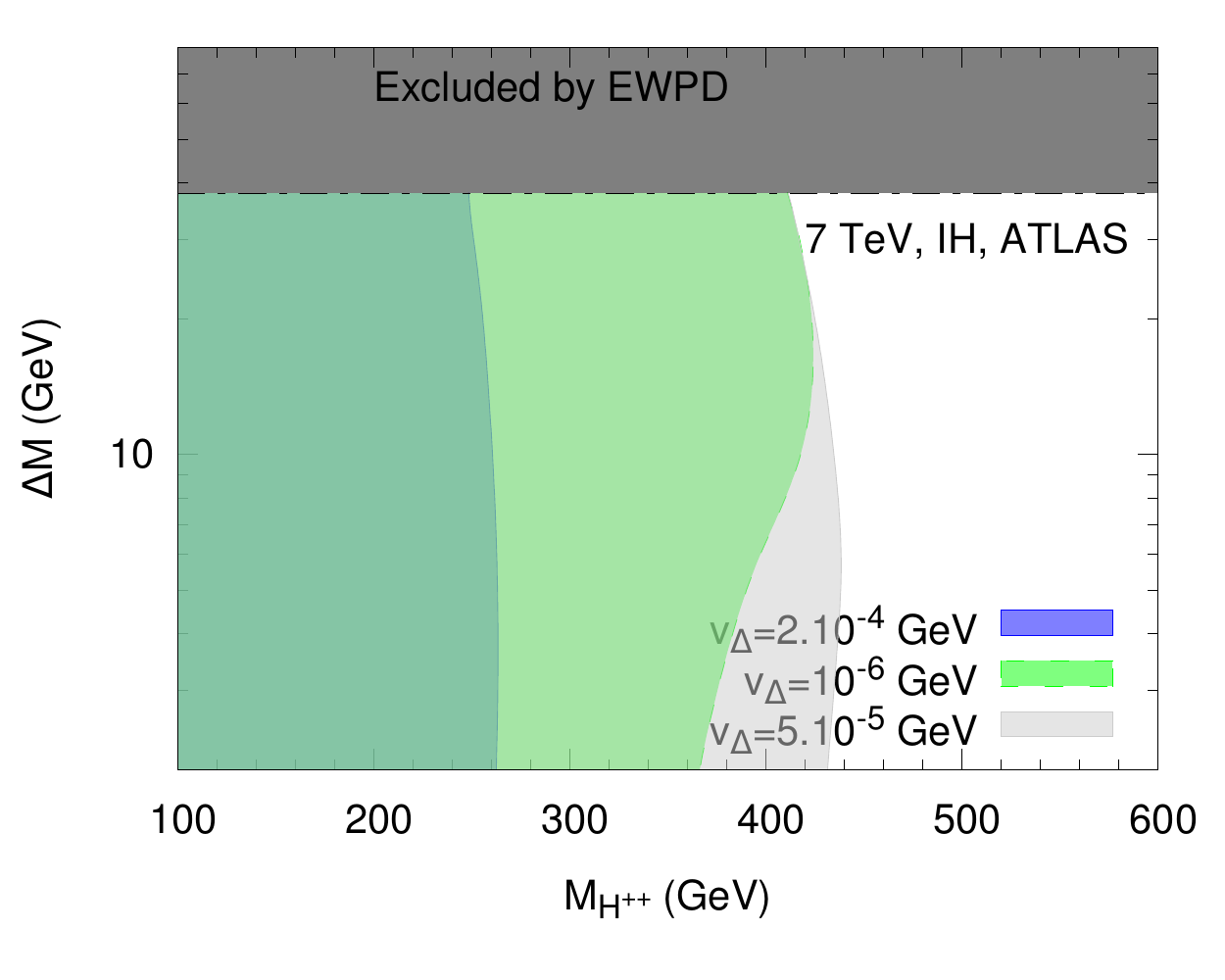}
\end{center}
\caption{ Exclusion region in the $M_{H^{++}}$--$\Delta M$ plane utilizing CMS (left) and
ATLAS (right) analyses for $v_\Delta=2\times10^{-4}$ GeV, $10^{-6}$ GeV and $5\times10^{-5}$ GeV.
For the upper (lower) panels, the NH (IH) neutrino  mass pattern is assumed.
\label{excl-ih}
\label{excl-nh}}
\end{figure}

In Fig.~\ref{excl-ih}, we plot exclusion region in the $M_{^{++}}$--$\Delta M$ plane obtained by
including all the processes contributing to $H^{++}$--$H^{--}$ pair-production for NH and IH in Eq.~(\ref{BRij}).
In the left panel, we utilize the four-lepton analysis as performed by CMS and
in the right panel, the same-sign dilepton analysis of ATLAS has been utilized to constrain the parameter space
in the $M_{^{++}}$--$\Delta M$ plane.
For our analysis, we consider three values of triplet vacuum expectation
value, namely, $v_\Delta = 10^{-6}$ GeV, $5\times 10^{-5}$ GeV and $2\times 10^{-4}$ GeV.
We find that the parameter space is the most constrained for $v_\Delta=5\times 10^{-5}$ GeV and the least for $v_\Delta=2\times 10^{-4}$ GeV. Notice  that the constraints are weaker for NH as
BF($H^{++}\to e^+e^+ + \mu^+\mu^++e^+\mu^+$) is considerably
smaller than that for the case of IH as can be seen from the table (\ref{BRij}).

From Fig. \ref{BF-Hpp}, one finds that
BF($H^{\pm\pm}\to\ell^\pm\ell^\pm$) is around 15\%-40\% for $v_\Delta=2\times 10^{-4}$ GeV
when $M_{H^{++}}$ is 200-500 GeV
while for $v_\Delta=5\times 10^{-5}$ GeV and smaller, it is over 80\%.
Thus, the constraints on the doubly-charged boson mass gets stronger for smaller $v_\Delta$.
Note that in Figs.~\ref{excl-ih} there appears a peculiar behavior for $v_\Delta=10^{-6}$ GeV.
When mass splitting $\Delta M$ is very small, the bound on charged Higgs mass is very loose
while for relatively large $\Delta M>10$ GeV, constraints become comparable to the case of
$v_\Delta=5\times 10^{-5}$ GeV. This behavior has to do with the branching fraction of, e.g.,
$H^+ \to H^{++} W^+$  shown in Fig. \ref{BF-Hp}.
One can see that
BF($H^\pm\to H^{\pm\pm}W^{\mp*}$) is always below 90\% for $v_\Delta=10^{-6}$ GeV
unless $\Delta M>10$ GeV, and thus none of the processes for triplet production mentioned
above will contribute to pair-production of $H^{++}$ unless $\Delta M>10$ GeV.
On the other hand, for $v_\Delta=5\times 10^{-5}$ GeV,
the BF is more than 90\% even for $\Delta M$ as low as 2 GeV and thus have large number of
events for $H^{++}$--$H^{--}$ production which lead to stringent constraints on $M_{H^{++}}$
even for small $\Delta M$.

The gray region in $M_{H^{++}}$--$\Delta M$ plane for $\Delta M>38$ GeV is excluded by considering
electroweak precision constraints on $\lambda_5$, hence on $\Delta M$ \cite{Chun:2012jw}.
This bound on $\Delta M$ is found to be independent of doubly-charged boson mass $M_{H^{++}}$.
One can also see that bounds obtained by
utilizing ATLAS analysis are  stronger than those obtained by following CMS. This is because
ATLAS collaboration have considered same-sign di-lepton signals coming from the decay of
only one doubly-charged boson in pair production while CMS have looked at four lepton final states. 
It is clear that ATLAS would have large number of signal events as compared to CMS.

\section{SS4L signals at LHC8/13}

Apart from the well-studied same-sign di-lepton signals, there can appear also a novel
phenomenon of same-sign tetra-leptons indicating the neutral triplet--antitriplet oscillation \cite{Chun:2012zu}.
Such a signal would be an indisputable evidence for the discovery of a doubly-charged boson
in type II seesaw. For this to occur, one needs a condition for the oscillation parameter:
\begin{equation} \label{xis}
x\equiv {\delta M_{HA} \over \Gamma_{\Delta^0} }\gtrsim 1
\end{equation}
where $\delta M_{HA}$ is
the mass splitting (\ref{MHA}) between two real degrees of freedom of
the neutral triplet boson, and
$\Gamma_{\Delta^0}\simeq \Gamma(\Delta^0 \to H^+ W^{-*})$.
Arising
from the lepton number violating effect, $\delta M_{HA}$ is proportional to
$\xi^2$ and thus can be
comparable to the decay rate of $\Gamma_{\Delta^0} \approx G_F^2 \Delta M^5/\pi^3$ which is  also
quite suppressed for a small mass gap $\Delta M \approx M_{H^0} - M_{H^+}$.
Once the oscillation parameter is determined, one can calculate the production cross-sections for
the same-sign tetra-lepton final states from the following formula \cite{Chun:2012zu}:
\begin{eqnarray}\label{ss4l}
\sigma\left(4\ell^\pm + nW^{\mp^*}\right)&=&\nonumber \Bigg\{\sigma\left(pp\to
H^\pm \Delta^{0(\dagger)} \right)  \left[{ x^2\over 2( 1+ x^2)}\right]
 \mbox{BF}(\Delta^{0(\dagger)}\to H^\pm W^{\mp^*})\\\nonumber
&+&\sigma\left(pp\to
\Delta^0 \Delta^{0\dagger} \right)  \left[{2+x^2 \over 2(1+x^2)}
{ x^2\over 2(1+ x^2)} \right]
 \left[\mbox{BF}(\Delta^{0(\dagger)} \to H^\pm W^{\mp^*})\right]^2\Bigg\} \nonumber\\
&\times& \left[\mbox{BF}(H^\pm\to H^{\pm\pm} W^{\mp^*})\right]^2
\left[\mbox{BF}(H^{\pm\pm}\to \ell_i^\pm\ell_j^{\pm})\right]^2 .
\label{4l3W}
\end{eqnarray}
To analyse the effect of oscillation, let us define,
\begin{eqnarray}\label{chiB}
\chi_B& \equiv&
\left[{ x^2\over 2( 1+ x^2)}\right]
\mbox{BF}(\Delta^{0(\dagger)}\to H^\pm W^{\mp^*})
\left[\mbox{BF}(H^\pm\to H^{\pm\pm} W^{\mp^*})\right]^2
\left[\mbox{BF}(H^{\pm\pm}\to \ell_i^\pm\ell_j^{\pm})\right]^2 .
\label{4l3W}
\end{eqnarray}
which
determines the viability of SS4L signal originating from process $pp\to H^+\Delta^{0\dagger}$
at the LHC. It includes factors such as BF($\Delta^{0(\dagger)}\to H^\pm W^{\mp^*}$),
BF($H^\pm\to H^{\pm\pm} W^{\mp^*}$) and oscillation probability which are indispensable components for the occurrence of SS4L signal at the LHC.

\begin{figure}[t]
\begin{center}
\includegraphics[width=90mm]{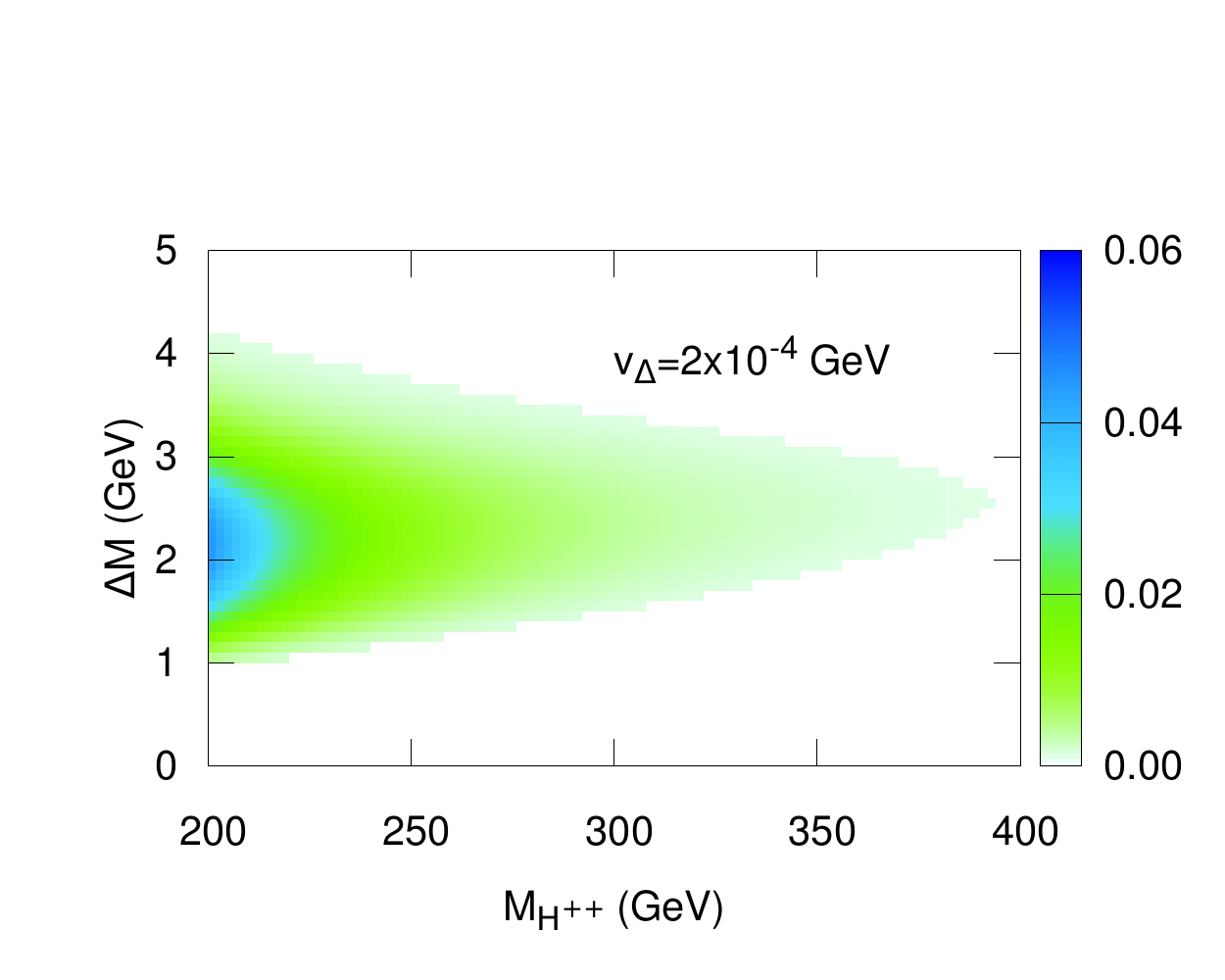}\hspace{-15pt}
\end{center}
\caption{The quantity $\chi_B$ (\ref{chiB}) in the $M_{H^{++}}$--$\Delta M$ plane for $v_\Delta=2\times10^{-4}$ GeV.
\label{fig:fac}}
\end{figure}

In Fig.~\ref{fig:fac}, we plot $\chi_B$ in the plane of $M_{H^{++}}$--$\Delta M$.
One can see that $\chi_B$ is sizable only in the limited range of
$\Delta M=(1,4)$ GeV at $M_{H^{++}}=200$ GeV.
The blue region in the figure is the area where probability of SS4L signal is maximum.
Lower values of $\Delta M$ is disfavored because BF($\Delta^{0(\dagger)}\to H^\pm W^{\mp^*}$) and
BF($H^\pm\to H^{\pm\pm} W^{\mp^*}$) are too suppressed (as seen from Fig.~\ref{BF-Hp})
while higher values are suppressed due to increase in $\Gamma_{\Delta^0}$ which leads to very small oscillation probability ($x^2 \ll 1$).
For larger $M_{H^{++}}$, $\Gamma_{\Delta^0}$ also increases and thus leads to narrowing down of allowed parameter space in the $M_{H^{++}}$--$\Delta M$ plane.

Let us now discuss if observable same-sign tetra-lepton signals can be obtained in the allowed parameter region of Fig.~\ref{excl-ih}.
For this analysis, we consider two values of $v_\Delta = 5\times 10^{-5}$ GeV and $2\times 10^{-4}$ GeV for which $M_{H^{++}}$ larger than about 400 and 200 GeV is allowed respectively,
and discard $v_\Delta=10^{-6}$ GeV which gives a vanishingly small oscillation probability $\chi_B$. We consider all triplet production processes which
can contribute to SS4L signal at 8 TeV (LHC8) and 13 TeV (LHC13) of LHC with 20 fb$^{-1}$ and 100 fb$^{-1}$ of integrated luminosities respectively.

\begin{figure}[t]
\begin{center} 
\includegraphics[width=90mm]{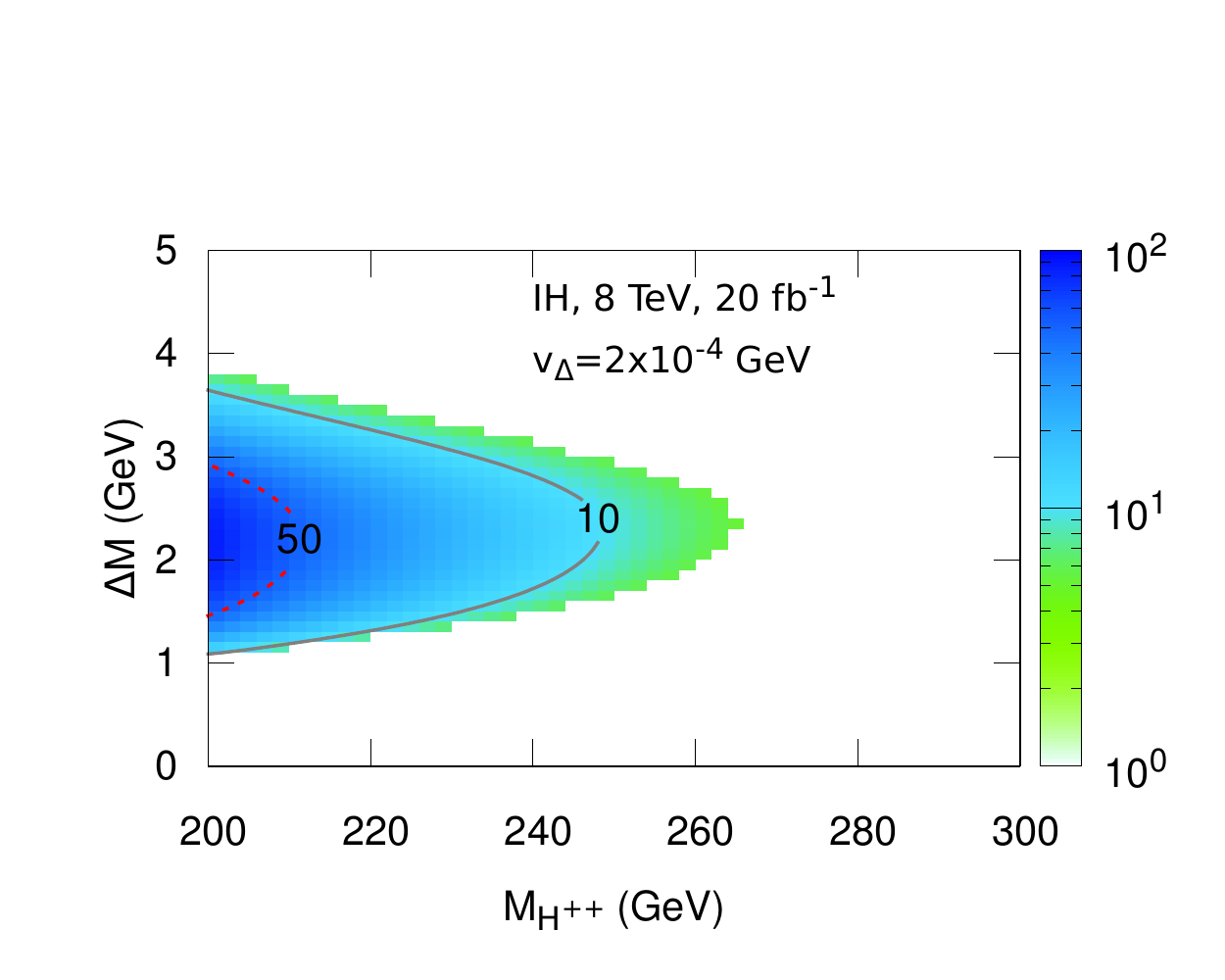}
\end{center}
\caption{Contour plots for SS4L signal numbers in the $M_{H^{++}}$--$\Delta M$ plane at LHC8  for $v_\Delta=2\times10^{-4}$ GeV.
Here the IH neutrino mass structure is taken.
\label{fig:8ss4l}}
\end{figure}

In the Fig.~\ref{fig:8ss4l}, we plot the number of SS4L events achievable for $v_\Delta=2\times 10^{-4}$ GeV at LHC8 with 20 fb$^{-1}$ of the integrated luminosity assuming the IH neutrino 
mass structure.  The signal numbers are smaller for NH.
The number of events for $v_\Delta=5\times 10^{-5}$ GeV at LHC8 are very low
for $M_{H^{++}}>400$ GeV, and thus this case is not interesting.
We find that a sizable number of SS4L events can be obtained in the range of $\Delta M \sim (1,4)$
GeV for which the oscillation probability is large enough. In order to see SS4L events at LHC8,
we need $M_{H^{++}}\lesssim 260$ GeV which, however,  is almost ruled out 
by the current ATLAS results shown in the lower left panel of Fig.~\ref{excl-ih}.

\begin{figure}[t]
\begin{center}\hspace{-38pt}
\includegraphics[width=90mm]{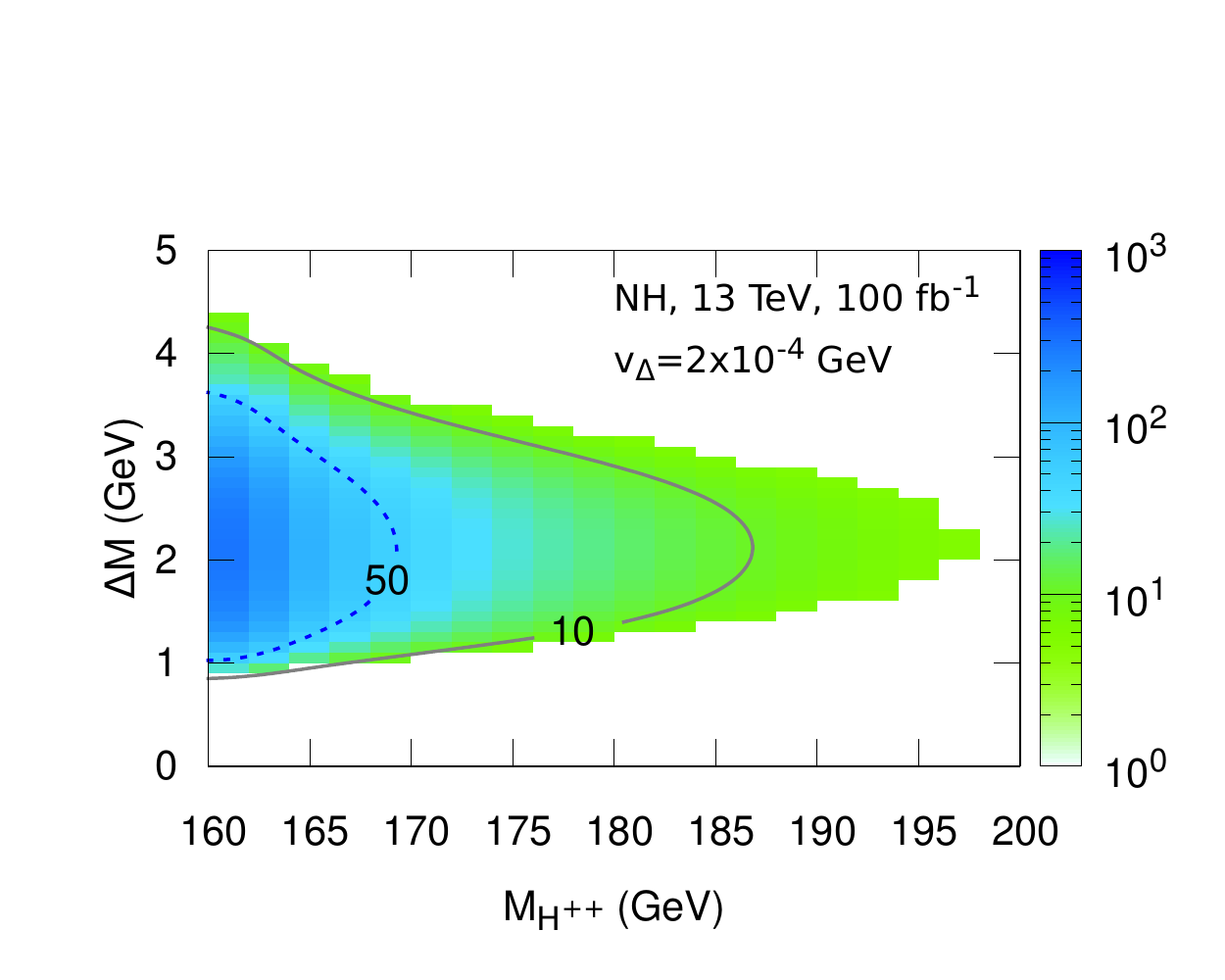}\hspace{-16pt}
\includegraphics[width=90mm]{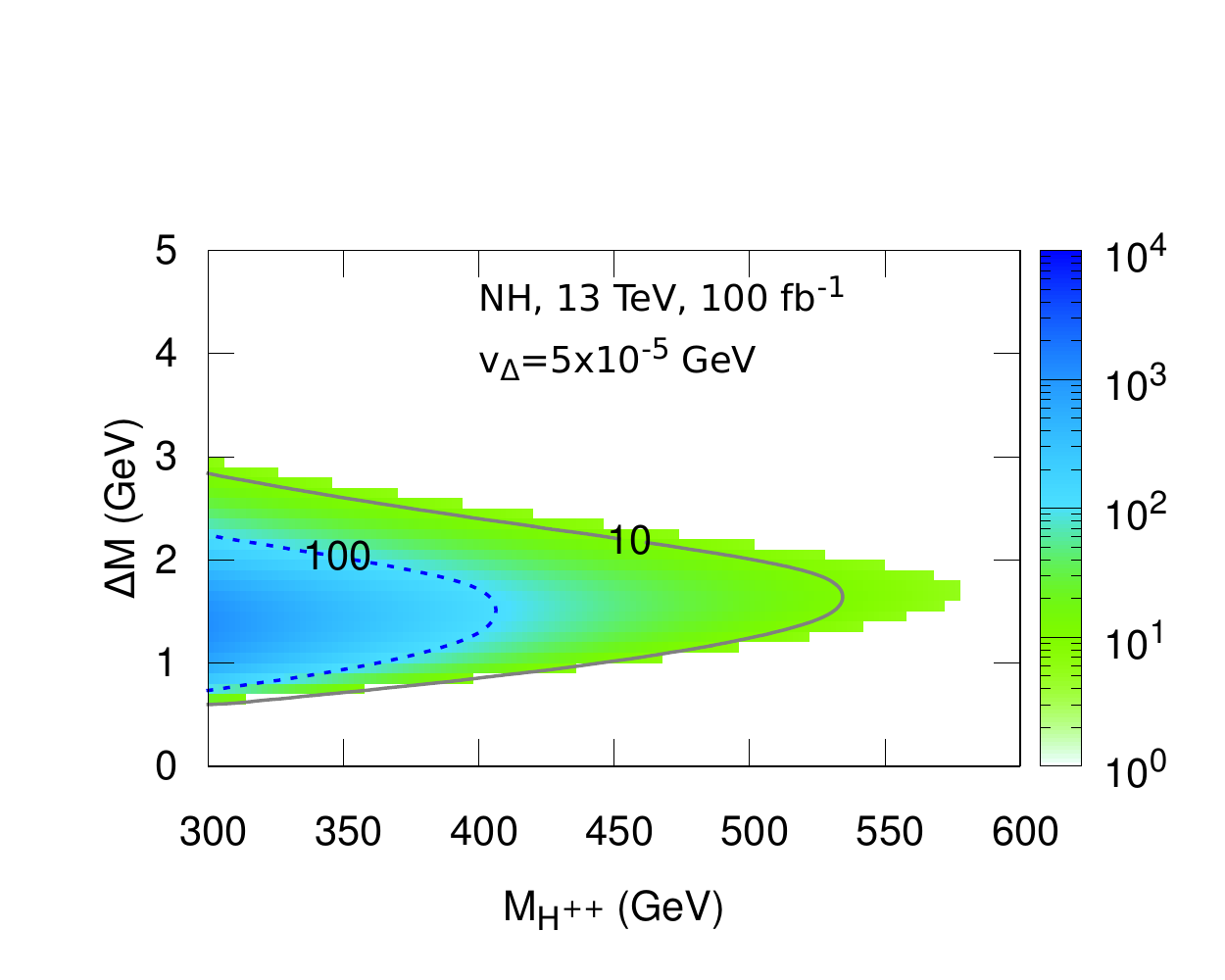}\\
\vspace{-10pt}
\hspace{-38pt}
\includegraphics[width=90mm]{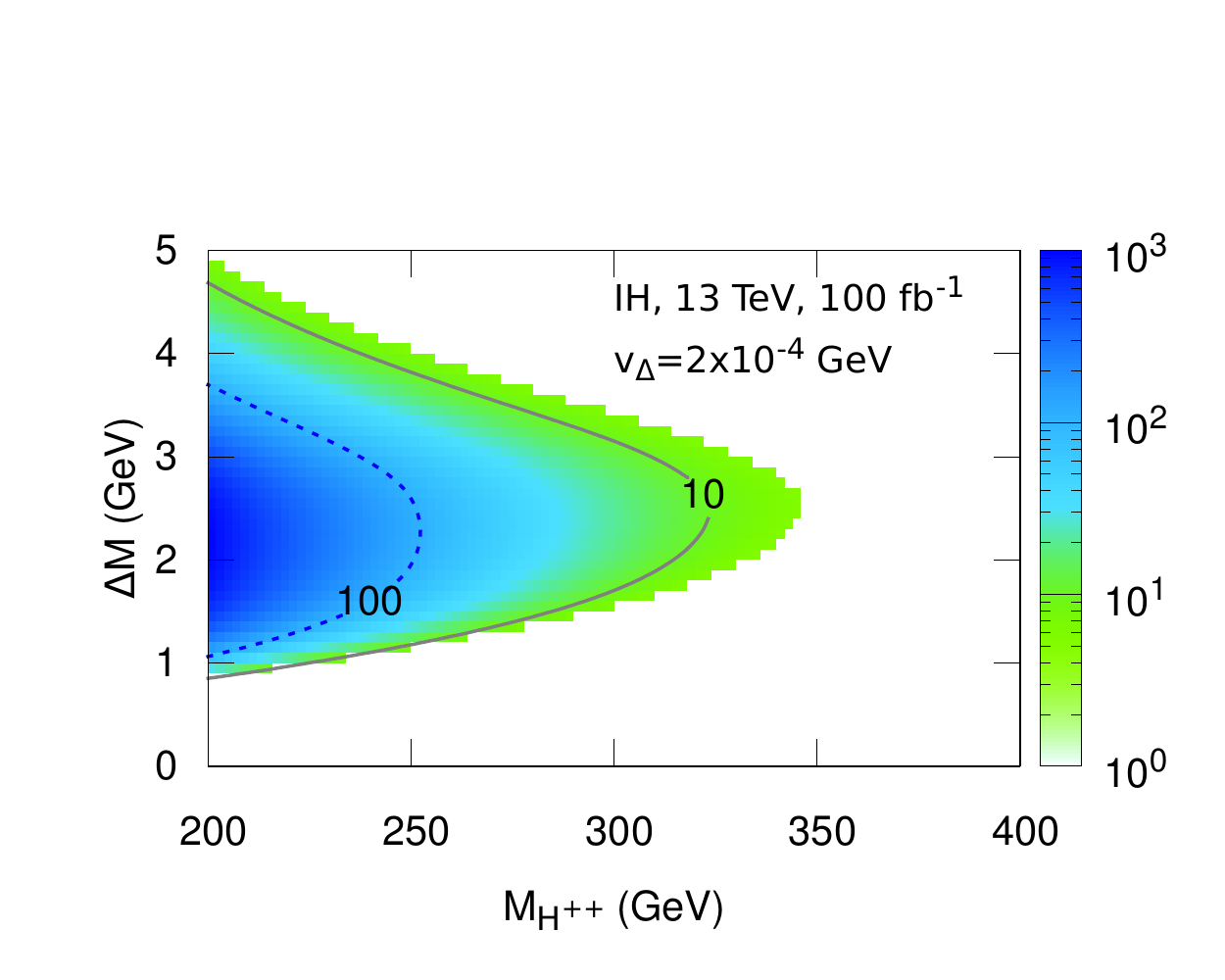}\hspace{-16pt}
\includegraphics[width=90mm]{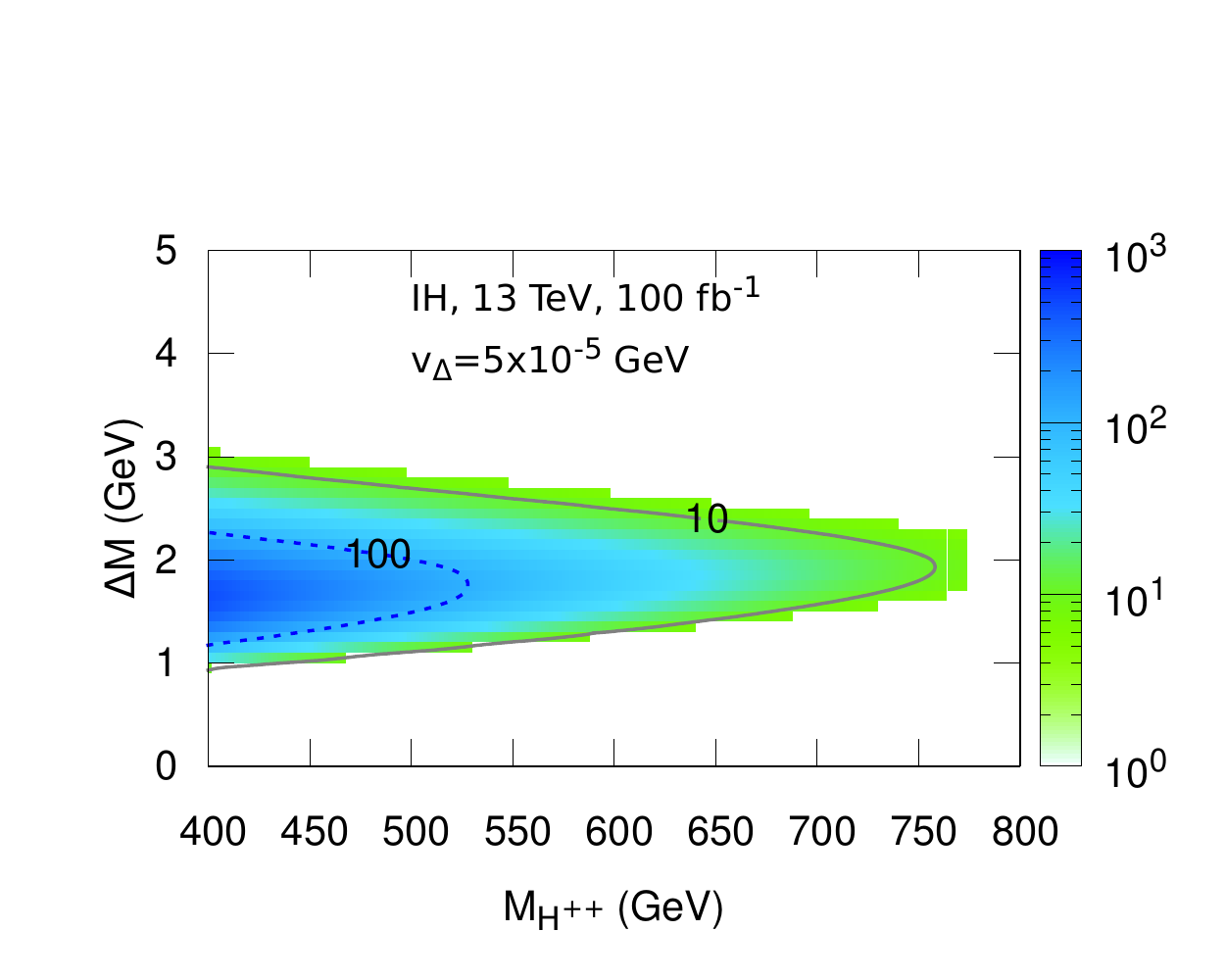}
\end{center}
\caption{ Contour plots for SS4L signal numbers  in  the $M_{H^{++}}$--$\Delta M$ plane at
LHC13 for $v_\Delta=2\times10^{-4}$ GeV and  $5\times10^{-5}$ GeV. The upper (lower) panels assume NH (IH) for
the neutrino mass pattern.
\label{fig:13ss4l}}
\end{figure}

Fig.~\ref{fig:13ss4l} shows the number of SS4L events at LHC13 with 100 fb$^{-1}$ of
the integrated luminosity for  $v_\Delta=2\times 10^{-4}$ GeV (left) and
$v_\Delta=5\times 10^{-5}$ GeV (right) taking NH (upper) and IH (lower)
for the neutrino mass structure.  As expected from the table (\ref{BRij}), more leptonic final states are obtained for IH
and thus better sensitivities for SS4L events are obtained. 
If we assume that 10 SS4L events would be sufficient for the claim of
$H^{++}$ discovery, then for $v_\Delta =2\times 10^{-4}$ GeV, $H^{++}$ can be probed up to 330 GeV at LHC13 in the case of IH. On the other hand, for $v_\Delta=5\times 10^{-5}$ GeV, $H^{++}$ can be probed up to 750 GeV at LHC13.
In the case of NH with $v_\Delta 2 \times 10^{-4}$ GeV, observale signals can be obtained only for $M_{H^{++}} < 200$ GeV 
which is exclued by the current ATLAS data, whereas $M_{H^{++}}$ up to 550 GeV can lead to observable SS4L signals  for $v_\Delta=5\times 10^{-5}$ GeV.

\section{Conclusion}

Type II seesaw model of neutrino mass generation introduces an $SU(2)_L$ triplet boson which
contains a doubly-charged scalar and thereby leads to peculiar collider signatures.
Collider phenomenology of the triplet boson sector depends on three parameters:
the mass gap $\Delta M = M_{H^\pm} - M_{H^{\pm\pm}} \approx M_{H^0/A^0} - M_{H^+}$ among the triplet
components $H^{\pm\pm}, H^\pm$ and $H^0/A^0$, the doubly-charged boson mass $M_{H^{\pm\pm}}$,
and the triplet vacuum expectation value $v_\Delta$ (or the leptonic Yukawa coupling
$f_{ij}$ of the triplet). Considering the case of $\Delta M>0$ for which the doubly-charged boson
is the lightest, we studied the LHC bounds on its mass depending on $\Delta M$
and $v_\Delta$ utilizing the current CMS and ATLAS search for the doubly-charged boson from same-sign
di-lepton (SS2L) resonances. In the range of $\Delta M \gtrsim 1$ GeV,
the gauge decays of the heavier triplet components end up with producing
doubly-charged bosons and associated $W^*$'s and thus augment the search limit of $M_{H^{\pm\pm}}$. 
On the other hand, the bound is weakened for larger $v_\Delta$ for which
the leptonic decay modes of the triplet bosons are more suppressed.
The results are summarized in Fig.~\ref{excl-ih} taking three representative values of $v_\Delta$
for the cases of two neutrino mass hierarchies (NH and IH).

When the tiny mass splitting between two neutral components $H^0$ and $A^0$ is
comparable to the decay rate $\Gamma_{H^0/A^0}$, there can appear an oscillation phenomenon
which leads to pair-production of same-sign doubly-charged bosons and thus
same-sign tetra-lepton (SS4L) final states at the LHC.
For allowed parameter region from the current SS2L search, we analyzed the prospects for
observing SS4L signals at LHC8 and LHC13 which are summarized in Figs.~\ref{fig:8ss4l} and \ref{fig:13ss4l}.
Note that more leptonic final states (with $e$ and $\mu$) are produced in the case of IH
compared to NH and thus better search sensitivity is obtained for IH.
Observable SS4L signature can be obtained only in the limited region of
$\Delta M \sim 1-4$ GeV and probed up to $M_{H^{\pm\pm}} \sim 750$ GeV at the LHC13
with 100 fb$^{-1}$ of the integrated luminosity for the most probable case of IH
with $v_\Delta = 5 \times 10^{-5}$ GeV. On the other hand,
the SS4L search becomes much more restricted
due to the reduced number of leptonic final states or smaller oscillation probability
for larger or smaller $v_\Delta$. In the case of $v_\Delta = 2 \times 10^{-4}$ GeV,
SS4L signals can be observable up to $M_{H^{\pm\pm}} \sim 350$ GeV.

\section*{Acknowledgments}
EJC was supported by SRC program of NRF Grant No.\ 2009-0083526
funded by the Korea government (MSIP) through Korea Neutrino Research Center.

\end{document}